\documentclass[aps,twocolumn,nofootinbib,pra,10pt,superscriptaddress]{revtex4-1}

\usepackage{amsmath,amsfonts,amssymb,amsthm,physics,graphicx}
\usepackage{comment}
\usepackage{caption}
\usepackage{subcaption}
\usepackage{bbm}

\usepackage{hyperref}
\hypersetup{
    colorlinks=true,
    linkcolor=blue,
    filecolor=blue,      
    urlcolor=blue,
    citecolor=blue,
    }

\DeclareMathOperator{\her}{Her}
\DeclareMathOperator{\pos}{Pos}
\DeclareMathOperator*{\argmax}{arg\,max}

\begin{document}
	
\title{Informationally Complete Orbital Angular Momentum Tomography with Intensity Measurements}

\author{M. Gil de Oliveira}
\author{A. L. S. Santos Junior}
\affiliation{Instituto de F\'{i}sica, Universidade Federal Fluminense, 24210-346 Niter\'{o}i, RJ, Brazil}
\author{P. M. R. Lima}
\affiliation{Departamento de Física, Universidade Federal de Minas Gerais, 31270-901 Belo Horizonte, MG, Brazil}
\author{A. C. Barbosa}
\author{B. Pinheiro da Silva}
\affiliation{Instituto de F\'{i}sica, Universidade Federal Fluminense, 24210-346 Niter\'{o}i, RJ, Brazil}
\author{S. Pádua}
\affiliation{Departamento de Física, Universidade Federal de Minas Gerais, 31270-901 Belo Horizonte, MG, Brazil}
\author{A. Z. Khoury}
\affiliation{Instituto de F\'{i}sica, Universidade Federal Fluminense, 24210-346 Niter\'{o}i, RJ, Brazil}

\date{\today}
	
\begin{abstract}
    In this work we study the tomography of the spatial structure of light. We develop a simple technique that allows one to perform the tomography over the space of fixed order modes. The technique is based on two spatially resolved intensity measurements, the second of which is performed after the light field has undergone an astigmatic transformation implemented by a tilted lens. We demonstrate that this method is informationally complete within the considered subspaces, which is experimentally verified both in the intense and the photocount regime. We also study the effect of obstructions in our ability to reconstruct the states. The work here presented is expected to help the development of characterization techniques in the field of structured light and in its application for both classical and quantum information protocols.    
\end{abstract}
	
\maketitle

\section{Introduction}

The field of structured light has found promising applications in areas such as optical communication \cite{PhysRevApplied.11.064058, Gibson:04, wang2012terabit}, optical tweezers \cite{PhysRevApplied.14.034069,PhysRevLett.131.163601, padgett2011tweezers} and quantum cryptography \cite{Mirhosseini_2015, Sit:18, PhysRevLett.113.060503, d2012complete, souza2008quantum}. Central to them is the transverse spatial structure of light, and, therefore, its characterization is extremely important, as reflected by the multitude of approaches that have been proposed in recent years. Methods in which a Spatial Light Modulator (SLM) is part of the measurement apparatus are commonly used \cite{hiekkamaki_near-perfect_2019, jia_vector_2023, PhysRevX.5.041006}. Machine learning techniques have also been employed \cite{da_silva_machine-learning_2021,PhysRevResearch.5.013142,PhysRevApplied.17.054019}. These techniques have been successfully applied to the classification of structures through atmospheric turbulence \cite{krenn_communication_2014,9888083}. In \cite{zhou_direct_2021}, a technique based on using the polarization to find the density matrix in position basis was also introduced while, in \cite{rambach_robust_2021}, self-guided tomography was used to perform accurate tomography in high dimensional spaces.

In this work, we perform the characterization using only spatially resolved intensity measurements over our light field, which are easily implemented by cameras. To achieve this, we explore a complete mathematical correspondence between this problem and the one of quantum state tomography \cite{paris_quantum_2004, heinosaari_mathematical_2011}. Such correspondence provides us with a versatile method, which has already been explored in earlier works \cite{jezek_reconstruction_2004}.

This method is particularly simple to implement experimentally, can be performed fast and imposes no power loss during the measurement stage, being only restricted by the efficiency of the detectors. This stands in contrast with SLM based tomography methods, in which each passage through the SLM incurs at considerable losses, which is undesirable, specially when working with very faint sources. The SLM also has a slow refresh rate, which combined with the fact that the tomography of a single state may require the measurement of many projections, makes it so that the tomography may take a long time to be performed. Although the use of Digital Micromirror Devices (DMDs) may alleviate the speed issue, they are subject to even higher losses.

Nonetheless, intensity based methods suffer from two important issues: the first of them is shared with any tomographic method, which is the fact that the spatial structure of light has infinitely many degrees of freedom, and would need an infinite number of measurements to be fully characterized. This is dealt with by assuming that the relevant information to be retrieved is restricted to a finite dimensional subspace. The second issue, which is particular of intensity based tomography, is that it is insensitive to the phase of the field, at least in a direct manner. This issue was circumvented in \cite{PhysRevLett.103.250402} by considering a subspace formed by Laguerre-Gaussian modes of positive topological charge. Although the phase was not directly measured, it was possible to infer it from the intensity, provided that one can be assured that the light field is within the subspace.

In the first part of the work we will explore a method that allows one to perform the tomography in more general subspaces, which includes Laguerre-Gaussian modes of both positive and negative topological charge, for which a single intensity measurement is insufficient.This technique is based on a second intensity measurement, performed after the light field has undergone an astigmatic transformation implemented by a tilted lens \cite{BEIJERSBERGEN1993, vaity2013measuring, buono_eigenmodes_2022, PhysRevLett.72.1137}. Although it was already introduced in \cite{da_silva_machine-learning_2021}, its application was restricted to pure states and the theory behind it was absent, as the tomography in the reference was based on machine learning. Here we explicitly demonstrate that this method is informationally complete within the subspace of fixed order modes, which is experimentally verified both in the intense and the photocount regimes. In the latter, we also show how the quality of the reconstruction depends on the number of detections.

In the second part of the work, we deal with a related problem: how can one perform the tomography of the spatial structure of light in the presence of obstructions? This is important for free space communications using structured light, where the large distances involved combined with diffraction makes it so that the beam may arrive at the receiver larger than the detector. Based on the theoretical description, we experimentally demonstrate how one is still able to perform the tomography in the presence of obstructions, which here is implemented by a blade and an iris, which progressively block larger portions of the beam.

The paper is then organized as follows: in section \ref{sec:quantum_state_tomography} we review the theory of quantum state tomography. In section \ref{sec:tomography of the spatial structure of light}, we discuss how to frame the characterization of the spatial structure of light within this framework. In section \ref{sec:experiment intense}, we present our experiments in the intense regime, while in section \ref{sec:experiment photocount}, we present our experiments in the photocount regime. In section \ref{sec:obstructed measurements}, we study the effect of obstructions in our ability to reconstruct the states. Finally, in section \ref{sec:conclusion}, we draw our conclusions.

\section{Quantum State Tomography}
\label{sec:quantum_state_tomography}

We now briefly review the theory of quantum state tomography. Let $\her(\mathcal{H}) \subset \mathcal{H}$ be the set of Hermitian operators acting on a Hilbert space $\mathcal{H}$, whose dimension is $d$. The state of a quantum system is represented by an element $\rho$ of the set of positive semi-definite operators $\pos(\mathcal{H}) \subset \her(\mathcal{H})$ such that $\Tr \rho = 1$. Observables are represented by elements $A \in \her(\mathcal{H})$ so that its expectation value is given by $\expval{A} = \Tr \left(\rho A\right)$. A Positive Operator Valued Measure (POVM) is a set of observables $\{\Pi_m\} \subset \pos(\mathcal{H})$ with the property that $\sum_m \Pi_m = \mathbbm{1}$, where $\mathbbm{1}$ is the identity operator. These operators model the possible outcomes of an experiment: outcome $m$ happens with probability $p_m = \Tr \left(\Pi_m \rho \right)$. This is Born's rule. The POVM conditions ensure that $p_m \ge 0$ and $\sum_m p_m = 1$. The POVM formalism has been used for minimum quantum tomography of polarization states \cite{Ling2006}, qubits and qudits in path states \cite{Pimenta2010,Pimenta13,Cardoso2019} or in vector vortex beams \cite{Mustafa2022}. 

To solve the tomography problem, one first chooses an orthonormal set of traceless operators $\{\omega_1, \ldots, \omega_{d^2-1}\} \subset \her(\mathcal{H})$. For definiteness, we set the $\omega$'s as the generalized Gell-Mann matrices \cite{Bertlmann_2008}, which come in three types:
\begin{equation}
    \label{eq:Gell-Mann_matrices}
    \begin{aligned}
        X_{jk} &= \frac{\ketbra{u_j}{u_k} + \ketbra{u_k}{u_j}}{\sqrt{2}}, \\
        Y_{jk} &= \frac{i(\ketbra{u_j}{u_k} - \ketbra{u_k}{u_j})}{\sqrt{2}}, \\
        Z_j &= \frac{1}{\sqrt{j + j^2}} \left( \sum_{r=1}^j \ketbra{u_r} - j \ketbra{u_{j+1}} \right) .
    \end{aligned}
\end{equation}
In the above equations, $j$ assumes the values $1, \ldots d-1$, while $k$ assumes the values $j+1, \ldots, d$. Furthermore, $\{ \ket{u_1}, \ldots, \ket{u_d}\}$ is our chosen basis set of $\mathcal{H}$.

We can then specify an arbitrary state $\rho$ by a list of real coefficients $\boldsymbol{\theta} = (\theta_1,\ldots,\theta_{d^2-1})$ such that
\begin{equation}
    \label{eq:rho_as_func_of_theta}
    \rho = \rho(\boldsymbol{\theta}) = \frac{\mathbbm{1}}{d} + \sum_{n=1}^{d^2-1} \theta_n \omega_n.
\end{equation}
We call $\boldsymbol{\theta}$ a (generalized) Bloch vector. By taking the Hilbert-Schmidt inner product of equation \eqref{eq:rho_as_func_of_theta} with respect to the POVM elements, we arrive at
\begin{equation}
    \label{eq:linear_system_for_theta}
    \mathbf{q} = T \boldsymbol{\theta}.
\end{equation}
Here, $q_m = p_m - \Tr \Pi_m / d$ and $T$ is a matrix with entries $T_{mn} = \Tr \left(\Pi_m \omega_n \right)$. The rows of $T$ are simply the coefficients of the POVM elements in the basis $\{\omega_n\}$.

If $T$ is injective, the linear system of equations \eqref{eq:linear_system_for_theta} has a unique solution given by
\begin{equation}
    \label{eq:inversion_formula}
    \boldsymbol{\theta} = (T^\dagger T)^{-1} T^\dagger\boldsymbol{q}.
\end{equation}
In this case, the POVM is said to be informationally complete. Otherwise, if $T$ is not injective, the solution is no longer unique and the POVM is said informationally incomplete. In physical terms, this means that our measurements are not sufficient to uniquely determine the state, as there will be at least two states that give the same experimental statistics. 

One should observe, nonetheless, that the probabilities that figure in \eqref{eq:inversion_formula} are not directly measurable, and can only be \textit{estimated}. One simple estimation method is to substitute them by the observed experimental frequencies $\hat{p}_m = N_m / N$, where $N_m$ is the number of times that outcome $m$ was observed and $N = \sum_m N_m$ is the total number of observations. Then, by applying \eqref{eq:inversion_formula}, we get estimators $\hat{\boldsymbol{\theta}}_{LI}$ and $\hat{\rho}_{LI}$ for the coefficients $\boldsymbol{\theta}$ and for the state $\rho$. This is the so-called Linear Inversion Estimator, and is the simplest solution to the tomography problem. We utilize this estimator for a part of the work, but it is important to notice that it has some shortcomings: first, it may produce density matrices that are not positive semi definite. Therefore, we always project the output of this estimator to the closest valid density operator, utilizing the algorithm of Ref. \cite{smolin_efficient_2012}. It is also only reliable for many observations, otherwise the experimental frequencies may strongly deviate from the true probabilities. When necessary, we will discuss other estimators that circumvent these issues.

\section{Tomography of the spatial structure of light}
\label{sec:tomography of the spatial structure of light}

Consider a transversally structured light field and let
\begin{equation}
    \Gamma(\mathbf{r}, \mathbf{r}^\prime) = \expval{\mathcal{E}^\dagger(\mathbf{r}) \mathcal{E}(\mathbf{r}^\prime)}
\end{equation}
be its correlation function. Here, $\mathcal{E}(\mathbf{r})$ is the electric field operator at the point $\mathbf{r} = (x,y) \in \mathbb{R}^2$ in the detection plane. Moreover, let us define a normalized correlation function by
\begin{equation}
    \gamma(\mathbf{r}, \mathbf{r}^\prime) = \frac{\Gamma(\mathbf{r}, \mathbf{r}^\prime)}{\int_{\mathbb{R}^2} \Gamma(\mathbf{r}, \mathbf{r}) d\mathbf{r}} .
\end{equation}
Then, we may regard $p(\mathbf{r}) = \gamma(\mathbf{r}, \mathbf{r})$ as a probability density function. A detector, which covers a region $\mathcal{R}$, has the probability of making a detection given by \cite{loudon_quantum_2000}
\begin{equation}
    \label{eq:probability_of_detection}
    p(\mathcal{R}) = \int_{\mathcal{R}} p(\mathbf{r}) d\mathbf{r}.
\end{equation}

This situation can be described within the framework of quantum state tomography as follows \cite{jezek_reconstruction_2004}: we interpret $\gamma(\mathbf{r}, \mathbf{r}^\prime) = \mel{\mathbf{r}}{\rho}{\mathbf{r}^\prime}$ as the position representation of a quantum state $\rho$ in the Hilbert space of square integrable functions with the standard inner product. Here, $\ket{\mathbf{r}}$ is the eigenstate of the position operator with eigenvalue $\mathbf{r}$. Then, the expression for the detection probability \eqref{eq:probability_of_detection} is the expectation value of the projector 
\begin{equation}
    \Pi(\mathcal{R}) = \int_{\mathcal{R}} \ketbra{\mathbf{r}}{\mathbf{r}} d\mathbf{r}
\end{equation}
onto the functions with support in $\mathcal{R}$, i.e., $p(\mathcal{R}) = \Tr \left[\rho \Pi(\mathcal{R}) \right]$. The collection of such operators over a set of disjoint regions $\mathcal{R}_1, \ldots, \mathcal{R}_m$ such that $\bigcup_i \mathcal{R}_i = \mathbb{R}^2$ forms a POVM.

This POVM describes the measurements performed by cameras: the probability $p(\mathcal{R})$ can be experimentally estimated by the value read by the pixel on the camera, normalized by the sum of the values of all the pixels, given that enough detections are obtained. These are the probabilities that figure in the linear inversion formula \eqref{eq:inversion_formula}, and that are subsequently used to estimate the Bloch vector $\boldsymbol{\theta}$ and the density operator $\rho$.

It is clear that this POVM is not informationally complete in the space of the square integrable functions as it is finite, while the state $\rho$ is infinite dimensional. Nonetheless, it is often the case where the relevant information is restricted to a finite dimensional subspace. This is typical of a communication scenario where a basis $\{u_1({\mathbf{r}}), \ldots, u_d(\mathbf{r})\}$, previously agreed upon, generates an encoding space. The fact that a state belongs to this encoding space means that density operator can be expressed as
\begin{equation}
    \label{eq:rho_in_fixed_order}
    \rho = \sum_{j,k=1}^{d} \rho_{jk} \ketbra{u_j}{u_k}, \ \ \ \rho_{jk} = \mel{u_j}{\rho}{u_k},
\end{equation}
or, equivalently, the correlation function can be written as
\begin{equation}
    \label{eq:gamma_in_fixed_order}
    \gamma(\mathbf{r}, \mathbf{r}^\prime) = \sum_{j, k=1}^{d} \rho_{jk} u_j(\mathbf{r}) u_k^*(\mathbf{r}^\prime).
\end{equation}

The matrix elements of the POVM elements with respect the basis functions are given by
\begin{equation}
    \label{eq:povm_mel}
    \Pi(\mathcal{R})_{jk} = \mel{u_j}{\Pi(\mathcal{R})}{u_k} = \int_{\mathcal{R}} u_j^*(\mathbf{r}) u_k(\mathbf{r}) d\mathbf{r}.
\end{equation}

These can be used to write the entries of the matrix $T$, which are retrieved from the expressions
\begin{subequations}
    \begin{equation}
        \Tr \left[ \Pi(\mathcal{R}) X_{jk} \right] = \sqrt{2}  \Re \int_{\mathcal{R}}  u_j^*(\mathbf{r}) u_k(\mathbf{r}) d\mathbf{r};
    \end{equation}
    \begin{equation}
        \label{eq:trace_Yjk}
        \Tr \left[ \Pi(\mathcal{R}) Y_{jk} \right] = \sqrt{2}  \Im \int_{\mathcal{R}}  u_j^*(\mathbf{r}) u_k(\mathbf{r}) d\mathbf{r};
    \end{equation}
    \begin{equation}
         \begin{aligned}
            \Tr \left[ \Pi(\mathcal{R}) Z_{j} \right]  = \frac{1}{\sqrt{j + j^2}} \left( \sum_{r=1}^j \int_{\mathcal{R}} \abs{u_r(\mathbf{r})}^2 d \mathbf{r} \right. \\ \left. - j \int_{\mathcal{R}} \abs{u_{j+1}(\mathbf{r})}^2 d \mathbf{r} \right).
         \end{aligned}
    \end{equation}
\end{subequations}

Each column of $T$ corresponds to one of the above expressions, while each row corresponds to a particular pixel region $\mathcal{R}$. We then see that the choice of our basis functions and detection regions completely determines the matrix $T$. Apart from that, the estimated probabilities obtained from our measurement and the trace 
\begin{equation}
    \Tr \Pi(\mathcal{R}) = \sum_{j=1}^{d}\int_{\mathcal{R}} \abs{u_j(\mathbf{r})}^2 d\mathbf{r},
\end{equation}
completely determine the linear system of equations \eqref{eq:linear_system_for_theta}. The solution of this system gives us the coefficients $\boldsymbol{\theta}$, which can be used to reconstruct the density operator $\rho$ using \eqref{eq:rho_as_func_of_theta}.

\section{Informationally incompleteness for fixed order modes}
\label{sec:experiment intense}

We will use the Hermite-Gaussian modes
\begin{equation}
    \label{eq:HG_modes}
    HG_{mn}(\mathbf{r};w) = \mathcal{N}_{mn} H_m\left( \frac{\sqrt{2}x}{w} \right) H_n\left( \frac{\sqrt{2}y}{w} \right) e^{-r^2/w^2}.
\end{equation}
as basis functions for our encoding spaces. Here, $H_n$ is a Hermite polynomial, $w$ is the beam's waist and $\mathcal{N}_{mn}$ is a constant that ensures that the modes are normalized to $1$. More precisely, we will choose the encoding space as the subspace of fixed order, which is the parameter $N = m + n$. For a given order $N$, the corresponding encoding space has dimension $d = N + 1$ and is generated by the basis elements $u_n = HG_{n-1, N-n+1}$, $n = 1, \ldots d$.

This encoding space is particularly convenient to work with due to a few reasons: first, all the modes evolve with the same Gouy phase. Therefore, the evolution of the intensity pattern is simple, with only a change in size. This means that the tomography can be performed without the need to take into account the propagation distance. Second, we observe that the Hermite-Gaussian modes are parametrized by their waist as well as their center, which, in \eqref{eq:HG_modes}, is set to be the origin. The decomposition in equations \eqref{eq:rho_in_fixed_order} and \eqref{eq:gamma_in_fixed_order} is only finite for a particular choice of the center and the waist, which are not known at the time of the measurement. We show, in Appendix \ref{sec:center_and_waist}, that, for the subspace of fixed order, these parameters can be estimated \textit{independently} of the Bloch vector, which also greatly simplifies the tomographic procedure.

Nonetheless, the POVM formed by a single direct intensity measurement is still incomplete in this subspace, which can be seen in multiple ways. As an example, the Laguerre-Gaussian $\ket{LG_{0,\pm1}}$ with radial order $0$ and topological charge $\pm 1$ can be expressed as $\ket{LG_{0,\pm1}} = \frac{1}{\sqrt{2}} \left( \ket{HG_{10}} \pm i \ket{HG_{01}} \right)$. Therefore, they are part of the subspace of fixed order $N = 1$. However, it is well known that these two modes have the same intensity pattern, and, therefore, cannot be distinguished by a single intensity measurement. Furthermore, the subspace of fixed order $N$ contains all the Laguerre-Gaussian modes $\ket{LG_{p, l}}$ such that $N = 2p + \abs{l}$. In particular the space will contain the pairs $\ket{LG_{p, \pm l}}$, which, once again, have the same intensity pattern.

\begin{figure}
    \centering
    \includegraphics[width=0.5\textwidth]{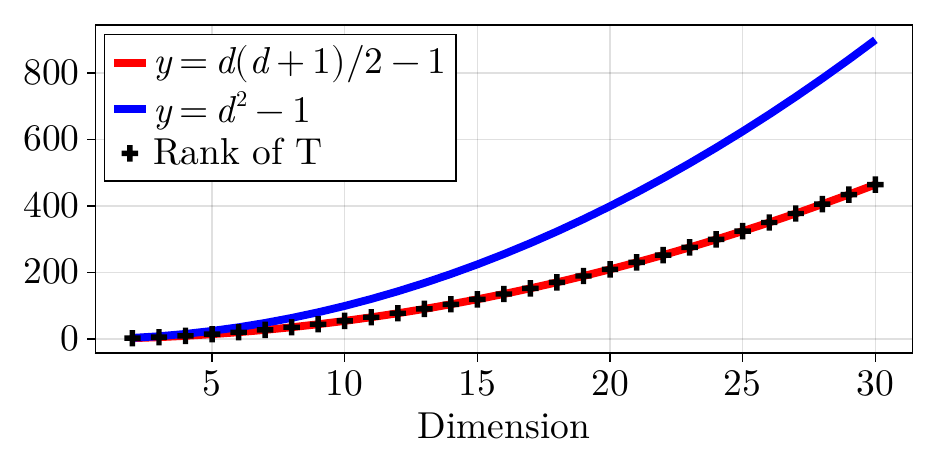}
    \caption{Rank of the matrix $T$ for the direct intensity measurement for different dimensions of the subspace of fixed order. We used a grid of $512 \times 512$ points, with a total size of $12$ arbitrary units in each direction. The modes had a waist of $1$ arbitrary unit and were centered at the origin.} 
    \label{fig:no_lens_rank}
\end{figure}

More generally, using the fact that the basis elements are real valued, one sees from \eqref{eq:gamma_in_fixed_order} that two states whose density matrices are the \textit{transpose} of each other will give the same intensity pattern. This can be quantified by observing that, according to \eqref{eq:trace_Yjk}, the POVM elements are orthogonal to all the $Y_{jk}$ operators. As there are $d(d-1)$ of them, the rank of the matrix $T$ can be at most $d(d+1) / 2 - 1$, which is smaller than the dimension of the space of density operators, $d^2 - 1$. This rank deficiency is illustrated in Fig. \ref{fig:no_lens_rank}.

\section{Informationally complete measurements with a tilted lens}

We now show that we can overcome this incompleteness by performing a second intensity measurement after the beam has undergone an astigmatic transformation. Such transformation is described by the unitary
\begin{equation}
\label{eq:mode_converter}
    U_\theta = \sum_{m,n=0}^{\infty} e^{i(m-n)\theta / 2} \ketbra{HG_{mn}}{HG_{mn}},
\end{equation}
which is physically realized by the use of cylindrical or tilted spherical lenses \cite{BEIJERSBERGEN1993,vaity2013measuring,buono_eigenmodes_2022}. The parameter $\theta$ can be adjusted experimentally by tuning the propagation distance and the tilt angle of the lens. More details on this calibration can be found on Appendix \ref{sec:calibration}. 

\begin{figure}
    \centering
    \includegraphics[width=0.5\textwidth]{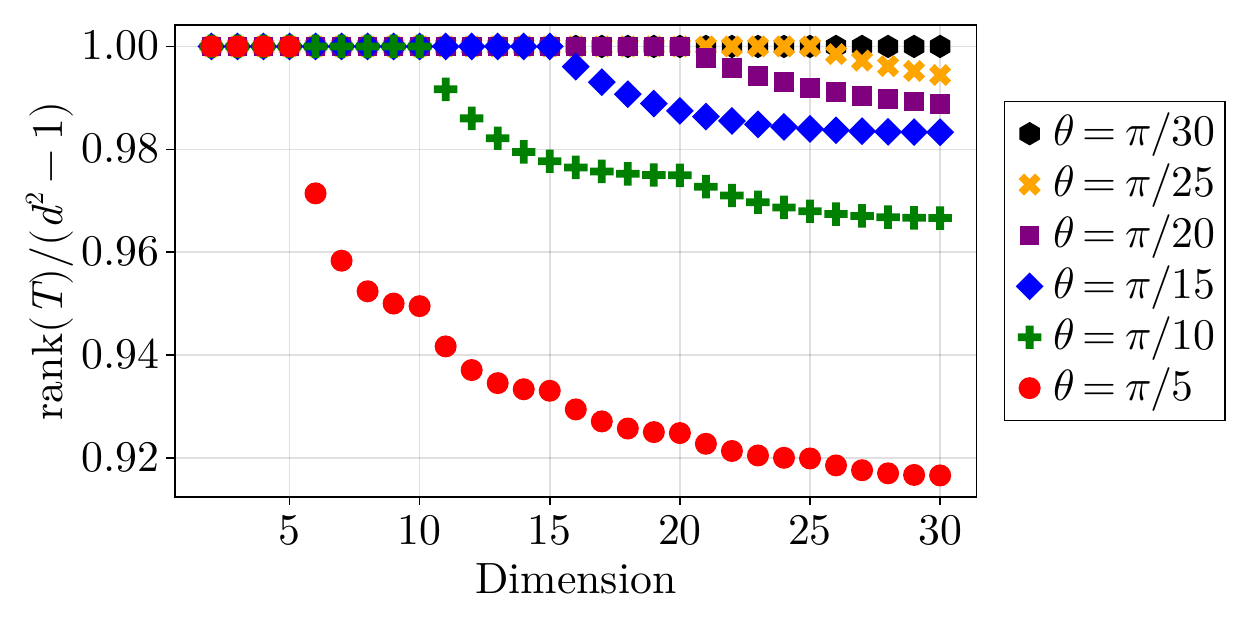}
    \caption{Normalized rank of the matrix $T$ corresponding to the two intensity measurements for different dimensions of the subspace of fixed order and different values for the parameter $\theta$. We used a grid of $512 \times 512$ points, with a total size of $12$ arbitrary units in each direction. The modes had a waist of $1$ arbitrary unit.} 
    \label{fig:lens_rank}
\end{figure}

To understand how this unitary is able to augment the POVM and make it informationally complete, let us denote $\Pi_\theta(\mathcal{R}) = U^\dagger_\theta \Pi(\mathcal{R}) U_\theta$. The addition of the converted image then generates the larger POVM $\{ \Pi(\mathcal{R})/2, \Pi_\theta(\mathcal{R})/2 \}$. For the new POVM elements, we have that
\begin{equation}
    \Pi_{\theta}(\mathcal{R})_{jk} = \mel{u_j}{\Pi(\mathcal{R})_\theta}{u_k} = e^{i(k-j)\theta}\int_{\mathcal{R}} u_j^*(\mathbf{r}) u_k(\mathbf{r}) d\mathbf{r},
\end{equation}
so that
\begin{equation}
    \Tr \left[\Pi_\theta(\mathcal{R}) Y_{jk}\right] = \sqrt{2}\sin \left[(k-j)\theta \right] \int_{\mathcal{R}} u_j^*(\mathbf{r}) u_k(\mathbf{r}) d\mathbf{r}.
\end{equation}
Therefore, we can avoid having $\Tr \left[ \Pi_\theta(\mathcal{R})Y_{jk} \right]= 0$ by properly choosing a value for $\theta$. Indeed, notice that, for a given dimension $d$, one has $\abs{k-j} < d - 1$. Therefore, by choosing $\theta = \pi / d$, we have that $\Tr \left[ \Pi_\theta(\mathcal{R})Y_{jk} \right] \neq 0$ for all $j,k$. This allows the POVM to be informationally complete, as confirmed by the numerical calculation shown in Fig. \ref{fig:lens_rank}, up to dimension $30$.

\begin{figure}
    \centering
    \includegraphics{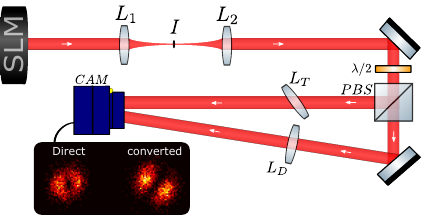}
    \caption{Experimental setup for tomography of the transverse structure of light. After reflection on the spatial light modulator (SLM), the beam is sent to a spatial filter. Then, it is split into two halves where direct and astigmatic imaging is performed.}
    \label{fig:setup}
\end{figure}

To verify the theory just described, we use the setup shown in Fig. \ref{fig:setup}: a beam of an intense HeNe laser at 633 nm wavelength is sent into a spatial light modulator (SLM). The SLM is used to synthesize the states of fixed order. After the SLM, the beam passes through the spatial filter formed by two spherical lenses, $L_1$ and $L_2\,$, with an iris ($I$) in between. After the spatial filtering, we send our beam through a half-wave plate ($\lambda / 2$) and a polarizing beam splitter (PBS), in order to control the intensity of each arm. Then, one arm is sent through a tilted spherical lens $(L_T)$, which implements our mode converter, while the other arm passes through a usual spherical lens ($L_D$). The mode converter is set to an angle $\theta = \pi/6$, as we work with dimensions up to $6$.

To synthesize the mixed states, we use the technique described in \cite{canas_evaluation_2022,Marques2015}, where a movie is played in the SLM, and the series of images are averaged in post-processing.

\begin{table}
    \centering
    \begin{tabular}{|c|c|c|}
      \hline
      \textbf{Dimension} & \textbf{Mean Fid. B.I. (\%)} & \textbf{Mean Fid. D.I. (\%)} \\\hline
      2 & 99.6 $\pm$ 0.2 & 94 $\pm$ 2 \\\hline
      3 & 97.4 $\pm$ 0.4 & 91 $\pm$ 2 \\\hline
      4 & 98.6 $\pm$ 0.2 & 89 $\pm$ 2 \\\hline
      5 & 98.1 $\pm$ 0.2 & 90 $\pm$ 2 \\\hline
      6 & 97.0 $\pm$ 0.3 & 90 $\pm$ 2 \\\hline
    \end{tabular}
    \caption{Mean fidelities between the reconstructed state and the desired one. B.I.: using both images. D.I.: using only the direct image. The error bars represent 95\% confidence intervals obtained through bootstrapping.}
    \label{tab:fidelities_intense}
\end{table}

In Table \ref{tab:fidelities_intense}, we assess the quality of our method using the fidelity $\mathcal{F}(\rho,\hat{\rho}) = \left(\Tr \sqrt{\rho \hat{\rho}} \right)^2$ between the desired state $\rho$ and the estimation $\hat{\rho}$ as a figure of merit. States belonging to the encoding space of fixed order are prepared. For each dimension, we perform the tomography of $100$ random mixed states sampled from the distribution described in \cite{zyczkowski_volume_1998} and calculate the average fidelity. The error bars for this result and for all the following ones represent 95\% confidence intervals obtained through bootstrapping \cite{chernick2011bootstrap}. We utilize $10^5$ samples. In the middle column, we show the fidelities when both the direct and the converted images are used to perform the tomography. We obtain high fidelities, even for dimension $6$, which evidence the quality of our approach. On the other hand, the right column shows the fidelities when only the direct image is used. We see that, in this case, the fidelities drop significantly, which experimentally confirms that the second measurement is indeed necessary to achieve informational completeness.

\section{Fixed Order Modes: Photocount Regime}
\label{sec:experiment photocount}

Up until here, we restricted ourselves to intense light beams that were measured by a standard CCD camera. This allowed us to use the simple Linear Inversion estimator. Now, we wish to turn to a regime where the light source is extremely faint, in such a way that one can now identify the individual photocounts. In particular, we wish to study how the quality of our estimate of the state behaves as we increase the observed number of photocounts. 

We can no longer utilize the Linear Inversion method because, now, there will not be a sufficient amount of observations to reliably estimate the probabilities of each outcome. This fact is taken into account by some tomographic techniques, such as Bayesian inference \cite{granade_practical_2016,lukens_practical_2020,landa_experimental_2022,lohani_demonstration_2023,blume-kohout_optimal_2010}, the variational tomography method \cite{Maciel2009, Maciel2011} or some of the methods based on neural networks \cite{PhysRevA.106.012409}. Here, we will employ the maximum likelihood estimator \cite{smolin_efficient_2012}
\begin{equation}
    \label{eq:ml_estimator}
    \hat{\rho}_{ML} = \argmax \mathcal{L}(N_1, \ldots, N_M | \rho)
\end{equation}
where the likelihood function is defined by
\begin{equation}
    \label{eq:likelihood_func}
     \begin{aligned}
         \mathcal{L}(N_1, \ldots, N_M | \rho) &= \prod_{m=1}^M p_m^{N_m} \\
         &= \prod_{m=1}^M \left[\Tr \left(\Pi_m \rho \right)\right]^{N_m}
     \end{aligned}
\end{equation}
and $N_m$ is the number of times that outcome $m$ was observed. In Eq. \eqref{eq:ml_estimator}, $\rho$ should be constrained to be a valid density operator. In order to perform the maximization, we utilized the Accelerated Projected Gradient (APG) with adaptive restart algorithm, implemented according to \cite{shang_superfast_2017}. As described in this reference, during each step of the optimization procedure, one checks if the state is still positive semi-definite, and, if not, the step is rejected and the algorithm proceeds with a smaller step size. This ensures that the output of the algorithm is always a valid density operator.

Due to experimental simplicity, we restrict ourselves to pure states. The tomography of these kinds of states with the aid of the tilted lens was demonstrated in \cite{da_silva_machine-learning_2021}, where the standard value $\theta = \pi/2$ was used. In that work, the tomography was performed over intense beams and using machine-learning, but there was no explanation of why the measurements performed do indeed carry enough information. The theory presented here explains why: let us consider a pure state $\ket{\psi} = \sum_j c_j \ket{u_j}$. The knowledge of the main diagonal elements $\rho_{jj}$ (determined from the coefficients corresponding to the $Z_j$) already allows the specification of $\abs{c_j}$, while the first secondary diagonal $\rho_{j,j+1} = \left[\Tr \left( \rho X_{j,j+1} \right) + i\Tr \left(\rho Y_{j,j+1}\right) \right] / \sqrt{2}$ allows the determination of the relative phases between each successive $c_j$, which completely characterizes the state. Then, the direct measurement gives us access to $\rho_{jj}$ and  $\Re \rho_{j, j+1}$, while the measurement preceded by the astigmatic transformation allows us to determine $\Im \rho_{j, j+1}$. The possibility of recovering \textit{pure} states from measurements is called unique determinedness \cite{PhysRevA.109.022425}. Note, nonetheless, that $\Tr \left[\Pi_{\pi/2}(\mathcal{R}) Y_{jk} \right]$ will still be zero when $k-j$ is even, so that this augmented POVM is still informationally incomplete, except in the case of order $N=1$, when the only possible value for $k-j$ is $1$.

\begin{figure}
    \centering
    \includegraphics[width=\linewidth]{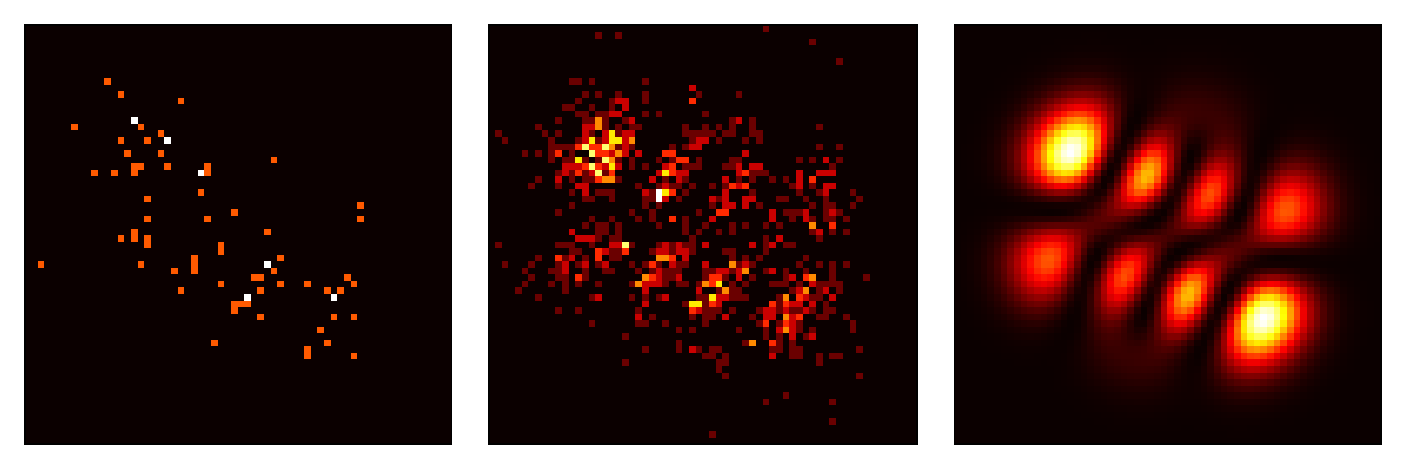}
    \caption{Evolution of the images as one increases the photocounts for a state of order $4$. The left and middle images are experimental results with $128$ and $2048$ photocounts. The right image is a simulation of the corresponding intensity distribution.}
    \label{fig:photocount_evolution}
\end{figure}

In order to test our method in the photocount regime, we use a similar experimental setup as shown in Fig. \ref{fig:setup}, the only differences being the use of an ICCD camera, which is sensitive to individual photocounts, and a filter, in order to not saturate the measurement. We also use the same basis as the previous experiment. Some experimental images are shown in Fig. \ref{fig:photocount_evolution}.

\begin{figure}
    \centering
    \includegraphics[width=\linewidth]{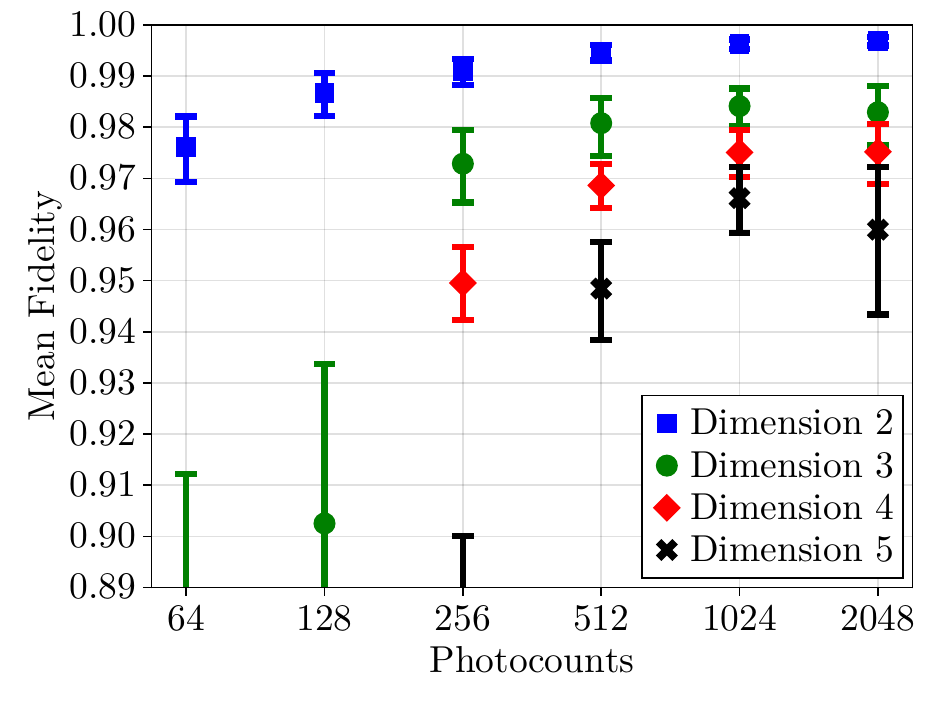}
    \caption{Mean fidelity as a function of the photocounts for different orders.}
    \label{fig:fidelities_photocount}
\end{figure}

We sample $50$ random states on which we perform the tomography. As we are only dealing with pure states, they are now sampled from the Haar measure \cite{mele2024introduction}, and, as explained, we are allowed to use a mode converter with parameter $\theta = \pi/2$. We also take as our tomographic estimate the eigenvector of the maximum likelihood estimator corresponding to the maximum eigenvalue.

The fidelities as a function of the photocounts for different orders are show in Fig. \ref{fig:fidelities_photocount}. As one expects, the fidelities rise as one increases the number of photocounts, reaching around $96\%$ for order $4$ at 2048 counts.

\section{Obstructed Measurements}
\label{sec:obstructed measurements}

In a scenario of free space communication using structured light, the beam may propagate through large distances, and, as a consequence, diffract. It can then become appreciably larger than the detection apparatus, which is a limiting factor for such an application. It is therefore important to understand the effect of obstructions in our ability to perform the tomography with intensity measurements.

In order to understand how the theory presented here has to be modified to deal with this problem, we first observe that, throughout this work, it was assumed that the measurements made formed a POVM. This is no longer true when the beam is partially blocked. In this case, it will be useful to define the region $\mathcal{R}_{D} = \cup_m \mathcal{R}_m$ covered by all the detectors and the operator
\begin{equation}
    \label{eq:g_position}
    g = \sum_{m=1}^{M} \Pi(\mathcal{R}_m) = \int_{\mathcal{R}_D} \ketbra{\mathbf{r}}{\mathbf{r}}d\mathbf{r},
\end{equation}
which is no longer approximately the identity. The  set $\{\Pi(\mathcal{R}_m)\}$ then fails to fulfill the POVM conditions.

One must then recognize that, in this setting, the experiment only allows us to estimate conditional probabilities $p_{m | D}$ of a detection at region $\mathcal{R}_m$ given that there was a detection at all. This is because the total detection probability $p_D = \Tr \left[\rho \Pi(\mathcal{R}_D) \right] =  \Tr(\rho g)$ is no longer $1$. To take that into account, we use the definition of conditional probabilities to conclude that
\begin{equation}
    \label{eq:conditional_probabilities}
    p_{m | D} = \frac{p_{m \cap D}}{p_D} = \frac{p_m}{p_D} = \frac{\Tr \left[\rho \Pi(\mathcal{R}_m) \right]}{\Tr \left(\rho g \right)}.
\end{equation}
The second equality comes from the fact that the joint probability $p_{m \cap D}$ satisfies $p_{m \cap D} = p_m$, because $\mathcal{R}_m \subset \mathcal{R}_D$. Of course, when $g = \mathbbm{1}$, equation \eqref{eq:conditional_probabilities} reduces to Born's rule.

To further advance our analysis, we also observe that a count in the detection area $\mathcal{R}_D$ is measurement in itself, which, according to the measurement postulate, updates the state through the non-unitary map
\begin{equation}
    \label{eq:density_operator_transform}
    \rho\mapsto \tilde{\rho} = \frac{A \rho A^\dagger}{\Tr \left(A \rho A^\dagger\right)} = \frac{A \rho A^\dagger}{\Tr \left(\rho g\right)}.
\end{equation}
Here $A$ is any operator that satisfies $A^\dagger A = g$. For definiteness, we obtain $A$ from the Cholesky decomposition of $g$ \cite{golub2013matrix}. If we also define a new set of operators
\begin{equation}
    \tilde{\Pi}(\mathcal{R}_m) = (A^{-1})^\dagger \Pi(\mathcal{R}_m) A^{-1},
\end{equation}
we have that
\begin{equation}
    \label{eq:conditional_probabilities_transformed}
    p_{m | D} = \Tr \left[\tilde{\rho} \tilde{\Pi}(\mathcal{R}_m) \right],
\end{equation}
which resembles the standard Born's rule. Noticing that the operators $\tilde{\Pi}_1, \ldots, \tilde{\Pi}_M$ form a POVM, because
\begin{equation}
    \sum_{m=1}^{M} \tilde{\Pi}(\mathcal{R}_m) = (A^{-1})^\dagger g A^{-1} = \mathbbm{1},
\end{equation}
we can then see that we have reframed the problem as a tomography of the collapsed 
state given by \eqref{eq:density_operator_transform}. All the techniques described in the previous sections can be applied to this new POVM, and the state $\tilde{\rho}$ can be reconstructed from the experimental data through equation \eqref{eq:conditional_probabilities_transformed}. Finally, the original state $\rho$ is reconstructed by inverting relation \eqref{eq:density_operator_transform}, and one obtains
\begin{equation}
    \label{eq:inverse_density_operator_transform}
    \rho = \frac{A^{-1} \tilde{\rho} (A^{-1})^\dagger}{\Tr \left[A^{-1} \tilde{\rho} (A^{-1})^\dagger \right]}.
\end{equation}

\begin{figure*}
    \centering
    \includegraphics[width=\linewidth]{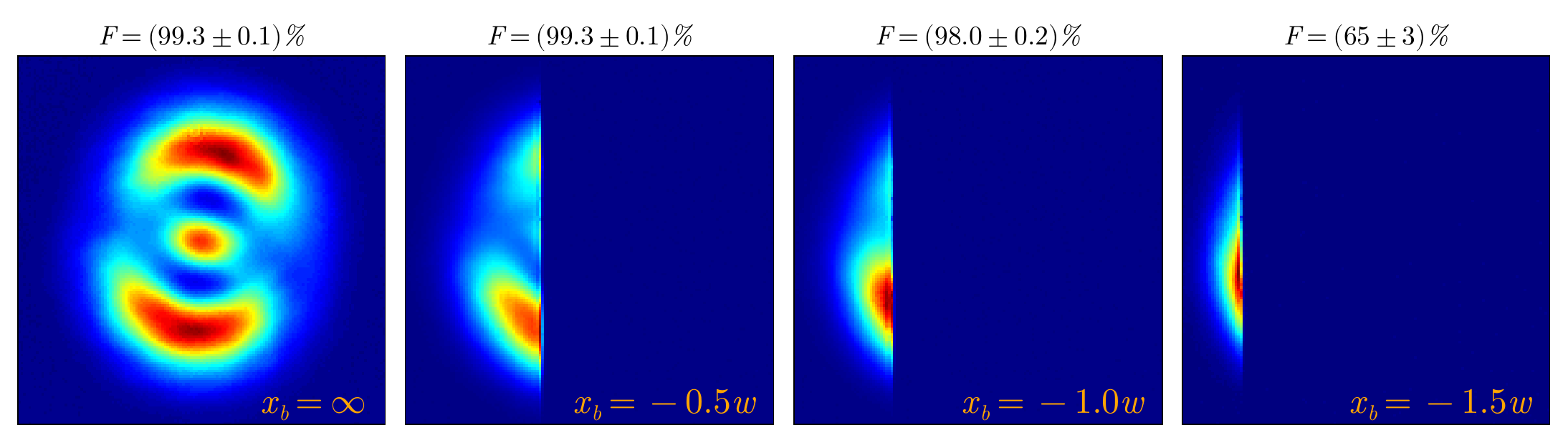}
    \caption{Example of a beam obstructed by a blade at different positions $x_b$. The blade positions are shown in the bottom right corner. On the top we display the mean fidelity for each obstruction.}
    \label{fig:blade_progression}
\end{figure*}

\begin{figure*}
    \centering
    \includegraphics[width=\linewidth]{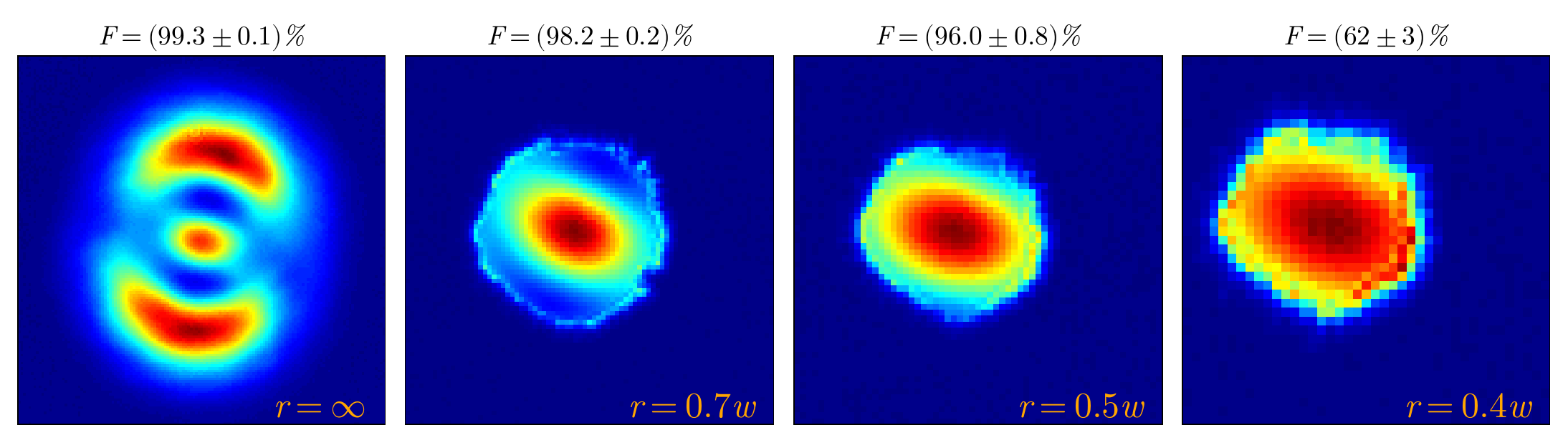}
    \caption{Example of a beam obstructed by an iris of different radii $r$. The radii are shown in the bottom right corner. On the top we display the mean fidelity for each obstruction.}
    \label{fig:iris_progression}
\end{figure*}

In order to study experimentally these ideas, we will work with a space spanned by the Laguerre-Gaussian modes 
\begin{equation}
    \label{eq:Lg}
    LG_{pl}(\mathbf{r};w) = \mathcal{N}_{pl} \left( \frac{\sqrt{2}r}{w} \right)^{\abs{l}} L_p^{\abs{l}} \left( \frac{2r^2}{w^2} \right) e^{-r^2/w^2 + il\phi}.
\end{equation}
Here, $\mathcal{N}_{pl}$ is a normalization constant, $L_p^{\abs{l}}$ is a generalized Laguerre polynomial, and $r$ and $\phi$ are the polar coordinates. Furthermore, $p$ is the radial order, while $l$ is the topological charge. For these modes, the order is given by $N = 2p + \abs{l}$. In particular, we will work in the $2$-dimensional subspace spanned by the modes $LG_{10}$ and $LG_{02}$, which are the modes of order $2$ with non-negative topological charge. In this subspace, a single direct image of the beam is informationally complete, which simplifies the experimental setup. We also go back to working with intense beams, allowing us to use simpler Linear Inversion estimator.

Instead of letting the beam propagate through large distances, we introduce physical obstructions in front of it, and image the plane of the obstruction into the camera using a pair of lenses. The obstructions considered will be a circular aperture, implemented by an iris, and a blade, which will progressively block larger portions of the beam.

We then prepare 100 different random mixed states, once again sampled from the distribution of reference \cite{zyczkowski_volume_1998}, and perform the tomography for each obstruction, measuring its quality using the fidelity. To specify the amount of obstruction, we assume a reference frame in which the beam is centered at the origin. The blade is then assumed to block every portion of the beam to the right of a position $x_b$. On the other hand, the iris is assumed to block everything outside a circle of radius $r$ centered at the origin. The results are shown in Figs. \ref{fig:blade_progression} and \ref{fig:iris_progression}. The images are experimental results of the intensity pattern of the same state under different obstructions. The values of the iris' radius and blade's position can be seen in the bottom right of the image, and are given in units of the beam's waist $w$. They are determined by first sending a Gaussian calibration beam and performing least squares fit to the intensity profile. At the top of the images, we show the mean fidelity over the 100 different states for each obstruction.

We see that the fidelity of the reconstruction decreases as the obstruction becomes larger, but that the method is still able to recover the state with a good accuracy until a certain threshold is hit. Afterwards, there is a sharp drop, and the mean fidelity reaches a value close to that of the mean fidelity between random states, at around $67 \%$ \cite{PhysRevA.71.032313}. This means that the tomography is no longer reliable. For the blade, this threshold is between $x_b = -1.0w$ and $x_b = -1.5w$, while, for the iris, it is between $r = 0.5w$ and $r = 0.4w$. This eventual drop is of course expected, but it is interesting to see that the method is able to recover the state even when the obstruction is seemingly large.

\section{Conclusion}
\label{sec:conclusion}

In this work, we developed a technique that allows one to perform the tomography of the spatial structure of light. We have demonstrated that the addition of a second intensity measurement, after the beam has undergone an astigmatic transformation, can make the POVM formed by intensity measurements informationally complete in the space of fixed order modes. This completeness is essential for reconstructing light structures 
involving positive and negative orbital angular momenta.

The theory was experimentally validated in the intense and photocount regimes, with high fidelities obtained in both cases. We have also shown how to deal with obstructions in the beam, and have demonstrated that one is still able to recover the state even when the obstruction is significant. This method will be useful for characterizing the transverse structure of light in a variety of applications, such as communication and information processing, both in the classical and quantum regimes.

\appendix

\section{Determination of center and waist}
\label{sec:center_and_waist}

Observe that the formulas \eqref{eq:HG_modes} and \eqref{eq:Lg} for the Hermite and Laguerre-Gaussian modes assume a frame of reference in which the beams are centered at its origin, but that usually will not be true when one performs actual measurements: the beam could be further displaced by a vector $\mathbf{r}_0 = (x_0,y_0)$. One must then find an estimator for this displacement, as well as for the beam's waist, in order to properly apply our tomography method.

This question has a straightforward solution if the beam is in a subspace of a known order. The order is the parameter $N = m+n$ for the Hermite-Gaussian modes and $N = 2p + \abs{l}$ for the Laguerre-Gaussian ones. In this case, one can estimate $x_0, y_0$ and $w$ independently of the Bloch vector, as we discuss as follows.

The Hermite and Laguerre-Gaussian modes can be described by an operator formalism, borrowed from quantum mechanics \cite{andrews2011structured}. We define $x, y$ as operators that act on a square integrable function $f$ through the formulas $(xf)(x,y) = x f(x,y)$ and $(y f)(x,y) = y f(x,y)$. In an analogous manner, we define $p_x = - i \partial_x$ and $p_{y} = -i \partial_y$. These operators follow the standard commutation relations $\left[ x, p_x \right] = \left[ y, p_y \right] = i$. We can then define the lowering operators
\begin{equation}
    a_x = \frac{x}{w} + \frac{i w p_x}{2}, \ \ \ a_y = \frac{y}{w} + \frac{i w p_y}{2},
\end{equation}
where $w$ is the waist of our modes.

The fundamental Gaussian mode $\ket{0} = HG_{00} = LG_{00}$ is the simultaneous solution of the eigenvalue equations $a_x \ket{0} = 0$ and $a_y \ket{0} = 0$. The Hermite-Gaussian modes can then be defined as
\begin{equation}
    \ket{HG_{mn}} = \frac{1}{\sqrt{m! n!}}\left( a^\dagger_x\right)^m \left( a^\dagger_y \right)^n \ket{0}.
\end{equation}

A field in a fixed order state may be expanded in the basis $u_n = HG_{n-1,N-n+1}(\mathbf{r}-\mathbf{r}_0; w), \ n=1,\ldots,N+1$ where $N$ is the order. Then, using the standard properties of the raising and lowering operators, one shows that, for any such field, the corresponding probability density function satisfies
\begin{subequations}
\begin{equation}
    \int_{\mathbb{R}^2} \mathbf{r} p(\mathbf{r}) d\mathbf{r} = \mathbf{r}_0,
\end{equation}
\begin{equation}
    \int_{\mathbb{R}^2} r^2 p(\mathbf{r}) d\mathbf{r} = r_0^2 +  \frac{(N+1) w^2}{2}.
\end{equation}
\end{subequations}

By performing measurements with a camera and recording the frequencies $\hat{p}_{mn}$ of photocounts in the pixel with coordinates $(m,n)$, one may easily estimate the desired parameters through the formulas
\begin{subequations}
\label{eq:center_and_waist}
\begin{equation}
    \hat{\mathbf{r}}_0 = \sum_{m,n} (m,n) \hat{p}_{mn},
\end{equation}
\begin{equation}
    \hat{w} = \sqrt{\frac{2}{N+1} \left( \sum_{m,n} (m^2 + n^2) \hat{p}_{mn} - \hat{r}_0^2 \right) }.
\end{equation}
\end{subequations}

\section{Calibration of the Tilted Lens}
\label{sec:calibration}

As presented in equation \eqref{eq:mode_converter} of the main text, the mode converter is implemented by a tilted lens. The angle $\theta$ is determined by the focal length of the lens, its tilt angle and the distance between the lens and the plane of measurement. Although a complete theory of this type of mode converter can be found in \cite{buono_eigenmodes_2022}, it is unnecessary for its practical implementation.

In order to calibrate the angle $\theta$ experimentally, we remember that this technique was originally introduced to measure the topological charge of a Laguerre-Gaussian mode \cite{vaity2013measuring}. In this case, the mode converter is tuned so that it corresponds to an angle $\theta= \pi / 2$ and implements the transformation \cite{BEIJERSBERGEN1993}
\begin{equation}
    \ket{LG_{mn}} \mapsto \ket{DHG_{mn}} = U_{\pi / 2} \ket{LG_{mn}}.
\end{equation} 
In this expression, $DHG_{mn}$ is a diagonal Hermite-Gaussian mode and $LG_{mn}$ is a Laguerre-Gaussian mode parametrized by $m$ and $n$. These values are related to the more common parametrization in terms of the topological charge $l = m - n$ and radial order $p = \min(m, n)$.

This tuning can be easily done experimentally because the diagonal Hermite-Gaussian modes are clearly recognizable in a camera. Therefore, one can, by trial and error, adjust the angle of the lens and the propagation distance until the image of the Laguerre-Gaussian mode is transformed into a diagonal Hermite-Gaussian mode. To implement the mode converter for a different parameter $\theta$, one simply uses $U_{-\theta}\ket{DHG_{mn}}$ as an input state, instead of the Laguerre-Gaussian mode, and then performs the same procedure.

\bibliography{refs.bib}	

\end{document}